# Speciation, ecological opportunity, and latitude

American Society of Naturalists 2013 Presidential Address

Dolph Schluter

Biodiversity Research Centre and Zoology Department
University of British Columbia
Vancouver, BC V6T 1Z4 Canada

**Abstract**

Evolutionary hypotheses to explain the greater numbers of species in the tropics than the temperate zone include greater age and area, higher temperature and metabolic rates, and greater ecological opportunity. These ideas make contrasting predictions about the relationship between speciation processes and latitude, which I elaborate and evaluate. Available data suggest that per capita speciation rates are currently highest in the temperate zone, and that diversification rates (speciation minus extinction) are similar between latitudes. In contrast, clades whose oldest analyzed dates precede the Eocene thermal maximum, when the extent of the tropics was much greater than today, tend to show highest speciation and diversification rates in the tropics. These findings are consistent with age and area, which is alone among hypotheses in predicting a time trend. Higher recent speciation rates in the temperate zone than the tropics suggest an additional response to high ecological opportunity associated with low species diversity. These broad patterns are compelling but provide limited insights into underlying mechanisms, arguing that studies of speciation processes along the latitudinal gradient will be vital. Using threespine stickleback in depauperate northern lakes as an example, I show how high ecological opportunity can lead to rapid speciation. The results support a role for ecological opportunity in speciation, but its importance in the evolution of the latitudinal gradient remains uncertain. I conclude that per-capita evolutionary rates are no longer higher in the tropics than the temperate zone. Nevertheless, the vast numbers of species that have already accumulated in the tropics ensure that total rate of species production remains highest there. Thus, "tropical evolutionary momentum" helps to perpetuate the steep latitudinal biodiversity gradient.

## Introduction

Despite great advances over the past two decades in our understanding of the origin of species (Coyne and Orr 2004; Grant and Grant 2008; Price 2008; Schluter 2009; Marie Curie Speciation Network 2012; Nosil 2012), the connection between speciation processes and global biodiversity patterns remains poorly understood (Butlin et al. 2009). Here, I investigate the relationship between speciation and the most conspicuous global pattern — the latitudinal gradient in species richness. In a large majority of higher-level taxa, many more species are found at tropical latitudes than in the temperate zone (Fischer 1960; Pianka 1966; MacArthur 1969; Gaston 2000; Willig et al. 2003; Hillebrand 2004). The pattern is evident across localities, biomes and regions in both the Old and the New World and in the northern and southern hemispheres. Speciation within latitudinal regions is the largest potential source of new species, and accordingly, a number of proposed explanations for the latitudinal diversity gradient are based on presumed variation in speciation processes (Mittelbach et al. 2007).

The possible roles played by speciation are framed by the dynamic history of the latitudinal richness gradient. The world was substantially hotter in the early Eocene (~55–48 Ma) (Zachos et al. 2001; Hansen et al. 2008), when tropical plant diversity was even higher than today (Jaramillo et al. 2006) and the latitudinal diversity gradient was more flat (Archibald et al. 2010; Mannion et al. 2014). Many taxa whose species are now restricted to the tropics ranged into the Eocene temperate zone, which was thermally less seasonal than the present time (Latham and Ricklefs 1993; Eldrett et al. 2009; Archibald et al. 2010) with little or no frost at latitudes as high as Antarctica (Pross et al. 2012). The world began to cool subsequently, with the first episodes of glaciation taking place about 34 Ma ago, accompanied by an increase in temperature seasonality (Crame 2001; Eldrett et al. 2009; Archibald et al. 2010). The mid-Miocene climatic optimum (17-15 Ma) also represented a world substantially warmer than today's (Zachos et al. 2001). A steep latitudinal diversity gradient developed as the earth continued to cool, during which the geographic extent of tropical taxa shrank toward the equator and the temperate zone shifted southward (Latham and Ricklefs 1993; Crame 2001; Pound et al. 2012). Thus, the modern large expanse of a seasonal temperate zone having significant frost days developed relatively recently. Additionally, episodic intrusions of ice sheets affected the temperate zone much more than the tropics, causing extinctions that steepened the gradient (Svenning 2003; Pushkina 2007; Sandel et al. 2011; Leslie et al. 2012; Jablonski et al. 2013; Villafaña and Rivadeneira 2014; Condamine and Hines 2015; Eiserhardt et al. 2015).

This history has led to the idea that the latitudinal gradient owes its steepness to the greater amount of time and area available to accumulate species in the tropics than the seasonal temperate zone (the *age and area* hypothesis) (Latham and Ricklefs 1993; Ricklefs 2004; Wiens and Donoghue 2004; Hawkins et al. 2006; Ricklefs 2006). Under this hypothesis, the imprint of historic tropical speciation on the modern diversity gradient persists to modern times despite areal contraction of the tropics and millions of years of seasonal temperate zone expansion. The reason typically emphasized is the presence of severe constraints on colonization of the temperate zone by tropical-adapted lineages. For example, tropical woody plants have rarely evolved tolerance to freezing, probably because of the susceptibility of water conduits to embolism (Tyree and Sperry 1989; Latham and Ricklefs 1993; Zanne et al. 2014). Nevertheless, a latitudinal gradient persists even in tropical plant lineages preadapted to freezing, such as those with narrow conduits or a non-woody life form (Moles et al. 2009; FitzJohn et al. 2014), and in marine invertebrates many of whose lineages have colonized temperate oceans (Jablonski et al. 2006). In any case, only a subset of tropical lineages can be expected to colonize the temperate zone, which means that this process will inevitably cause temperate species diversity to lag behind that of the tropics. In situ speciation has more potential to equalize the latitudinal gradient in the time available than colonization. Thus another explanation for history's imprint to the present day is that speciation has not yet made up the large initial shortfall.

Other views on the connection between the latitudinal diversity gradient and speciation processes (reviewed in Mittelbach et al. (2007)) can be grouped into two general hypotheses. One is the *temperature and metabolic rate* hypothesis, which attributes greater numbers of species in the tropics to faster genetic divergence between populations resulting from higher temperatures, higher metabolic rates, and shorter generation times. The other, which I will call the *ecological opportunity* hypothesis, attributes higher species diversity in the tropics to higher speciation rates owing to the greater number of ecological niches stemming from higher solar energy and annual productivity, reduced temperature seasonality, or stronger biotic interactions. Both alternative ideas attribute the latitudinal diversity gradient to ongoing mechanisms rather than historical legacy.

These three general hypotheses make different predictions regarding the relationship between latitude and the process of speciation. For example, they differ by whether higher speciation rates in the tropics are the result of faster evolution of reproductive isolation ("speed") or increased demographic rates governing the "metapopulation dynamics" of speciation, namely the frequency of establishment of spatially separated populations, the probability of population persistence, and the speed of establishment of secondary sympatry (Mayr 1963; Allmon 1992; Schluter 1998; Price 2008; Rabosky and Matute 2013; Dynesius and Jansson 2014; Price et al. 2014). The ecological opportunity hypothesis additionally addresses how the buildup of species diversity feeds back to affect rates of speciation subsequently. All three of these explanations for the latitudinal gradient assume that speciation is a rate-limiting step in the buildup of species diversity. Thus they assume that diversity has not everywhere reached an ecological limit or saturation point (Rohde 1992; Sax et al. 2007; Wiens 2011; Cornell 2013; Harmon and Harrison 2015).

Below I evaluate these hypotheses for the evolution of the latitudinal gradient. My focus here is on variation in speciation rather than colonization or extinction, which nevertheless also affect the gradient. I incorporate effects of species extinction indirectly. I treat episodic large-scale extinction that restarts the diversification clock, such as that caused by climate change and glaciation, as equivalent to reduced time/area for speciation. Episodic extinction is thus incorporated in the age and area hypothesis. Effects of chronic or baseline extinction, caused by ordinary fluctuations in population size, are considered when I compare trends in speciation with trends in diversification, which is the difference between speciation and species extinction.

Although comparative data on evolutionary rates allow tests of predictions of the three hypotheses, they provide limited insight into the mechanisms of speciation underlying latitudinal variation in rates. For example, comparative results presented below suggest that latitudinal patterns of speciation rate are associated with ecological opportunity. To evaluate this idea further, I use threespine stickleback to illustrate how the mechanisms by which niche availability leads to rapid speciation can be evaluated. I find evidence that ecological opportunity has both enhanced population persistence and generated divergent natural selection, which have contributed to rapid speciation in depauperate northern lakes.

## Predictions on speciation and latitude

By speciation I mean the evolution of reproductive isolation (Dobzhansky 1937; Mayr 1942). I assume that estimates of speciation rate obtained from the literature are based on this criterion or a reasonable surrogate. Variation in species definitions with latitude would complicate interpretation of the results summarized here. Speciation rate refers to the frequency with which new species are produced by a lineage, which is distinct from (though might be affected by) the rate of evolution of reproductive isolation. I assume that the definition of the tropics and temperate zone is consistent between data sources (e.g. Fine and Ree (2006)). Often, data on rates are provided as a function of latitude rather than as differences between zones, in which case I used regression to predict values at the equator (tropics) and at 40° latitude (temperate zone). Otherwise I refer to the tropics loosely as occurring between 30° S and 30° N, the approximate latitudes whose minimum annual temperature occasionally drops below freezing. I consider the temperate zone loosely as the latitudinal band between about 30° and 60°, N and S, which includes much of the boreal forest but not tundra or ice.

Evidence suggests that most speciation events probably involve some form of selection (natural and sexual), which can occur by either of two general mechanisms (Rundell and Price 2009; Schluter 2009; Unckless and Orr 2009). First, reproductive isolation might evolve as a consequence of divergent selection between contrasting ecological niches or environments (hereafter, ecological speciation). Second, separate populations might diverge by chance as they adapt to similar selection pressures, because they experience and fix different mutations that lead to reproductive incompatibilities (hereafter, mutation order speciation). Speciation resulting from the buildup of reproductive isolation by intragenomic conflict is an example of this second process. How either of these mechanisms should vary with latitude has been little addressed. If their relative importance differs according to alternative hypotheses for the gradient, then this could become an additional means to testing the hypotheses.

*Age and area*

Data suggest that speciation rate increases with increasing area (Losos and Schluter 2000; Kisel and Barraclough 2010). This process alone might explain why more species occur in the tropics, because large expanses of tropical habitats have been around for longer than similarly large expanses of seasonal temperate zone. Total land area is currently roughly similar between the tropics and the temperate zone, with most temperate lands concentrated in the northern hemisphere (Rosenzweig 1992; Chown et al. 2004). In contrast, surface area of the open ocean peaks between 20° N to 40° S and declines steeply toward the poles (Allen and Gillooly 2006). Area of shallow marine environments is highly unevenly distributed with latitude, being highest both in the tropics and at high latitude (Roy et al. 1998). Hence, under this hypothesis speciation rates today should be relatively similar between the tropics and temperate zone (except in the open ocean). However, rates are predicted to be highest in the tropics in the past.

Greater area likely affects speciation mainly by increasing the probability of population subdivision, and by increasing population size and thus persistence times (Losos and Schluter 2000; Gavrilets and Vose 2005; Kisel and Barraclough 2010). In other words, differences in speciation rate between large and small areas are predicted to result mainly from metapopulation dynamic processes rather than from effects of area on the rate of evolution of reproductive isolation. Large areas have greater environmental heterogeneity than small areas, but whether this increases the speed of reproductive isolation depends on the spatial scales of population subdivision and variation of environment. Area is expected to increase both mutation-order and ecological speciation, since both are affected by the dynamics of population subdivision and persistence.

Physical area matters less than "biologically effective area," measured in units of mean dispersal distance. Biologically effective area should be greater in the tropics if conditions there favor the evolution of narrower physiological tolerances (Janzen 1967; Ghalambor et al. 2006) and reduced dispersal (Cadena et al. 2012; Salisbury et al. 2012). In this case, speciation rates per unit physical area should be higher in the tropics compared with the temperate zone. If so, average physical distance between populations undergoing speciation should be reduced in the tropics relative to that in the temperate zone. This has been tested only by Lawson and Weir (2014), who instead found in birds that average geographic distance between the geographical range midpoints of diverging sister taxa was similar across latitudes (~ 1900 km).

*Temperature and metabolic rate*

Under the temperature and metabolic rate hypothesis, higher diversity in the tropics is caused by faster speciation resulting from the effect of higher temperature on mutation rates and generation times (Rohde 1992; Allen et al. 2006). Higher mutation rates are posited to result from greater DNA damage caused by higher free radical production at higher metabolic rates, and from the greater number of DNA replication errors accompanying more frequent generations (Rohde 1992; Gillooly et al. 2005). A higher supply rate of mutations would presumably generate more beneficial variants for natural selection to act upon (Rohde 1992; Bromham et al. 2015), leading to greater speed of evolution of reproductive isolation. The hypothesis explicitly assumes that speciation rate is mutation-limited. As a result, we might predict this mechanism mainly to affect rates of mutation-order speciation (Dowle et al. 2013), unless ecological speciation is also somehow mutation-limited. In principle, too high a mutation rate could slow mutation-order speciation by increasing the chances of the same advantageous mutations arising and fixing in separate populations (Unckless and Orr 2009).

The temperature and metabolic rate hypothesis applies mostly to ectotherms, and additional processes must be invoked to explain higher tropical diversity of endotherms. Indeed, basal metabolic rates in birds and mammals are slightly higher in the temperate zone than in the tropics (Lovegrove 2000; Wiersma et al. 2007) which under the temperature and metabolic rate hypothesis would predict a reverse latitudinal gradient. Likewise, there is little indication that endotherms have shortened generation times at tropical temperatures, but data are scarce. Longer activity periods result in faster development and more generations per year in tropical insects compared with temperate insects (Cardillo 1999; Archibald et al. 2010), but the opposite may be true in birds (Russell et al. 2004; Tarwater et al. 2011).

To date, little compelling evidence connects temperature to speciation or diversification rates via higher rates of mutation (Davies et al. 2004; Evans and Gaston 2005; Dowle et al. 2013; Bromham et al. 2015). Molecular substitution rates in ectotherms are indeed faster in the tropics than in the temperate zone (Wright et al. 2006; Wright et al. 2010; Bromham et al. 2015), though differences in endotherms are less clear (Bromham and Cardillo 2003; Gillman et al. 2009; Weir and Schluter 2011). Molecular evolution and diversification are often correlated, but the cause is not known (Davies et al. 2004; Bromham et al. 2015). Across a broad swath of animal phyla, transcriptome-wide amounts of synonymous and non-synonymous genetic variation within species are strongly correlated with life history characteristics of species but there is no trend with latitude (Romiguier et al. 2014). Rates of adaptive molecular evolution across latitudes, which are more relevant to speciation than rates of neutral divergence, have not been measured. Rates of phenotypic evolution, which are likely to be adaptive and often result in the evolution of some reproductive isolation as a by-product (Schluter 2009), have also rarely been compared with latitude. In birds, rates of divergence in body size between sister species and phylogroups are higher in the temperate zone rather than in the tropics, associated with higher rates of climate-niche divergence (Lawson and Weir 2014).

*Ecological opportunity*

Lastly, under the ecological opportunity hypothesis speciation rates should be highest in the tropics because of higher "ecological opportunity", defined as a diversity of evolutionarily accessible resources, such as food or enemy free space, underutilized by existing taxa (Schluter 2000; Losos 2010; Yoder et al. 2010). The concept is borrowed from the adaptive radiation literature, where ecological opportunity is often invoked to explain the bursts of speciation and ecological adaptation that occur in some lineages following colonization of a new or relatively depauperate environment, such as an oceanic archipelago or isolated lake or continent, following a mass extinction, or following the acquisition of a key evolutionary innovation (Lack 1947; Simpson 1953). The ecological opportunity hypothesis for the latitudinal gradient is that the tropics have more species than the temperate zone because they have a higher speciation rate resulting from a greater number of ecological niches. This is distinct from the view that more niches in the tropics allow more species to coexist there.

In an evolutionary interpretation, distinct niches can be equated with alternative fitness peaks in an adaptive landscape whose axes are phenotypic traits (Figure 1). Distinct peaks at a given geographical location represent different combinations of available resources or alternative combinations of traits that interact to determine high fitness on similar resources (Wainwright et al. 2005; Doebeli and Ispolatov 2010). Figure 1 illustrates a landscape at a single point in space, but the number, height and phenotypic position of peaks is expected to vary spatially. Adaptive peaks are typically visualized as static, but landscapes also vary temporally. Furthermore, ecological interactions between species generate frequency-dependent selection that alters the height and shifts the positions of peaks and may increase or decrease their number (Doebeli and Dieckmann 2000; Waxman and Gavrilets 2005; Rueffler et al. 2006).

The ecological opportunity hypothesis is based on the idea that the more underutilized adaptive peaks there are in a given area, the higher is the rate of speciation. One mechanism is by divergent selection between populations adapting to different peaks. For example, populations exploiting different food resources in different habitats would experience contrasting selection on traits, which may lead to rapid evolution of reproductive isolation as a by-product. Ecological opportunity can also increase the rate of speciation via the mutation-order process between spatially separated populations that exploit equivalent adaptive peaks. The effect can occur if the presence of a similar adaptive peak in two locations increases the success of establishment of spatially isolated populations, and the chances that they persist long enough to evolve reproductive isolation.

The ecological opportunity hypothesis might explain the known associations between species diversity and energy, annual productivity, precipitation and other environmental variables across latitudes (Currie 1991; Jetz and Fine 2012). Ecologists have long emphasized the role of such environmental variables in establishing limits to species coexistence. Under the ecological opportunity hypothesis defined here, environmental variables affect species diversity instead by affecting the speciation rate.

A related idea is the biotic interactions hypothesis, whereby a greater relative importance of biotic (relative to abiotic) agents of selection is posited to occur in the tropics (Dobzhansky 1950; Schemske 2009). This generates stronger and more frequent selection, increasing the rate of speciation in the tropics via faster evolution of reproductive isolation. For example, a greater influence of biotic interactions might generate greater spatial heterogeneity in selection pressures in the tropics, promoting speciation. Greater selection via biotic interactions in the tropics is also expected to result in more coevolution, possibly leading to a constantly changing optimum phenotype, again producing more divergent selection, continuous adaptation, and faster speciation (Schemske 2009). I consider this hypothesis to be a special case of the ecological opportunity hypothesis, because interactions with other species, whether as competitors, mutualists, hosts and parasites or predators and prey, represent niche axes that affect the number and phenotypic positions of adaptive peaks. Biotic interactions can lead to divergence and coevolution if, as is likely, they generate frequency dependent selection.

Are there fewer niches in the temperate zone than the tropics? This possibility is suggested by conspicuous latitudinal gradients in the diversity of ecological and life history phenotypes associated with latitudinal variation in environments. For example, the number of layers of forest vegetation declines with increasing latitude, which is thought to be a consequence of reduced incident solar radiation per unit of earth's surface area (Terborgh 1985). A narrower range of water conduit architectures in woody plants is feasible in the temperate zone than in the tropics because of the susceptibility of large-diameter conduits to embolism upon freezing (Zanne et al. 2014). Modeling of phenotypes on a climate grid suggests that more combinations of growth-related plant traits are viable in the tropics than in the temperate zone (Kleidon and Mooney 2000). The frequency of large-billed insectivorous bird species declines with latitude (Schoener 1971) in association with reduced biomass of large-bodied insects, itself the likely result of a shorter annual growing season (Schoener and Janzen 1968). Reduced seasonality of resources in the tropics likely also explains the many-fold higher number of specialized frugivorous bird species at 0° than at 40°N compared with only a 2 or 3 fold difference for bird species in general (Kissling et al. 2012). Herbivorous lizards and fish are largely restricted to warm climates, probably because vertebrates rely on gut endosymbionts to digest plant cell walls, which require high temperatures (Espinoza et al. 2004; Floeter et al. 2005). Not all phenotypes show a wider range in the tropics, however. The morphological diversity of passerine birds is similar between latitudes (Ricklefs 2012). The span of variation in tree height and specific leaf area are greater in the temperate zone than the tropics, whereas that for tree seed mass is similar between zones (Lamanna et al. 2014).

Biotic interactions are sometimes stronger in the tropics than in the temperate zone (Schemske et al. 2009), though whether this is a cause or a consequence of higher species diversity is not yet clear. Certainly many types of species interactions are present mainly or exclusively in the tropics (Schemske et al. 2009). However, there is little data on the effect of coevolution on speciation (Hembry et al. 2014). In phytophagous insects, rate of speciation is highest in association with shifts to novel hosts rather than by continuous association between host and phytophage (Hardy and Otto 2014), suggesting that new interactions are more potent than long-term coevolution in generating new species.

The major drawback of the ecological opportunity hypothesis is that ecological opportunity cannot be measured directly. Like cosmic dark energy, ecological opportunity can only be inferred from its effects. This difficulty makes the hypothesis challenging to test, and is also behind a lack of agreement on how increased species diversity should feed back on populations to affect ecological opportunity and speciation. If resources become used up as diversity increases, then ecological opportunity is expected to decline, predicting a slowdown in speciation. But since species are themselves resources, higher diversity might be expected to generate even more ecological opportunity, increasing speciation rate further (Whittaker 1977; Schemske 2002). Expectations likely differ depending on whether the effects are measured within or between trophic levels.

# Speciation, diversification, and latitude

Are speciation rates higher in the tropics than the temperate zone? Below I compare four types of data with latitude: diversification rates, speciation rates, divergence times of sister species, and rates of evolution of reproductive isolation. Estimates of speciation rates are likely to be highly uncertain, because they rest on models whose robustness to violations of assumptions is largely untested (Ricklefs 2007; Rabosky 2010; Rabosky and Goldberg 2015). It is not yet possible to contrast the effects of area, temperature, and ecological opportunity on ecological and mutation-order speciation across latitudes, which remains a future goal.

### *Diversification rates and latitude*

Estimates of diversification rates are not consistently lower in the temperate zone than in the tropics (Figure 2; Table A1). The median ratio of rates, temperate zone to tropics, is 0.96 ($n$ = 18) — which is scarcely a difference. Nevertheless, the estimates of diversification rate are highly heterogeneous, even when based on the same taxon (Fig. A1). The individual estimates of diversification rate, or their ratio, are based on any of a wide diversity of methods ranging from simple calculation of log species number divided by time to estimates based on fitting sophisticated trait-dependent birth-death models to phylogenetic trees. Most of the methods are summarized in Ricklefs (2007) and Rabosky and Goldberg (2015). Calculation details are described in Table A1. Data underlying new calculations from the original sources are deposited in the Dryad Digital Repository: http://dx.doi.org/10.5061/dryad.248vp (Schluter 2015).

Some of the variability among estimates of diversification ratio is accounted for by differences among studies in the time span covered by the investigation, as indicated by the oldest date analyzed (Figure 2; Spearman rank correlation between diversification rate and oldest date $r_S$ = −.55 ± 0.15 SE, $n$ = 18; Table A1; SE's are heuristic, because studies on the same taxon are unlikely to be independent). The younger the maximum date in an individual analysis, the more likely it is that estimated diversification is found to be similar in the temperate zone and the tropics (Figure 2). The older the maximum date, the more likely it is that diversification is found to be highest in the tropics. This relationship between diversification ratio and time span covered by analysis, if real, implies that relative diversification rates in the temperate zone and the tropics have changed over time. In particular, it suggests that diversification rates were historically higher in the tropics, but that rates are presently more similar.

*Speciation rates and latitude*

Speciation rate estimates show a median ratio, temperate to tropics, of 0.86, which is slightly lower than the diversification ratio but is calculated from a smaller sample size (Figure 3, $n = 11$). These estimates of speciation rate are based on a diversity of methods. Fossil-based estimates are based on the rate of new species occurrences. Most estimates, however, are based on birth-death models fitted to phylogenetic trees. Tree-based estimates of speciation rate are potentially biased because they are obtained against a background of unobserved extinctions. On the other hand, estimates of contemporary speciation rates are probably reasonably accurate, because not enough time has elapsed yet for many extinctions to have occurred.

As in the case of diversification rates, speciation rate estimates from published sources show a correlation with maximum date analyzed ($r_S = -.61 \pm 0.20$ SE. The older the maximum date, the more likely it is that estimated speciation rate in the temperate zone falls below that in the tropics. The younger the maximum date analyzed, the more likely it is that estimated speciation rate in the temperate zone *exceeds* that in the tropics. The trend, if real, implies that speciation rates were once highest in the tropics, and that the speciation gradient has become flattened and possibly even reversed in recent time, such that it is currently unrelated to, or opposite, the modern latitudinal diversity gradient.

*Sister species divergence times*

Divergence times ("ages") of sister species of birds and mammals are about a million years older at the equator than at 40° latitude, on average, the opposite of the pattern we might expect if speciation were fastest in the tropics (Weir and Schluter 2007; Weir and Price 2011). Mean age of splits between populations within species follows a similar latitudinal gradient (Weir and Schluter 2007; Weir and Price 2011; Botero et al. 2014). Stevens (2006) saw the same pattern in New World leaf-nosed bat species between 0 and 25° latitude. Perhaps the trend is a taxonomic artifact of greater numbers of species yet to be described in the tropics (Tobias et al. 2008). However, the pattern in birds holds even if restricted to sympatric species or to pairs forming a hybrid zone, which are those species most easily recognized (Weir and Price 2011). Additionally, in birds and mammals the latitudinal gradient in ages of sister species holds even within the temperate zone of North America north of 30°N, where species are well sampled. The same latitudinal age trend is seen for North American sister species of fish (Bernatchez and Wilson 1998; April et al. 2013). Unfortunately, comparisons for fish don't yet include tropical or Old World lineages such as the African cichlids, a unique hyper-diverse clade with many exceptionally young sister species (Seehausen 2006). Pyron and Burbrink (2009) found a flat relationship between age and latitude in 10 sister pairs of New World Lampropeltini snakes, whose peak in species richness straddles the temperate zone and the tropics between 19°N and 37°N. In birds within tropical latitudes of the New World, the youngest sister species tend to occur at higher elevation rather than in the tropical lowlands (Weir 2006). These patterns are inconsistent with the idea that speciation rates are currently highest in the tropics, at least in these taxa.

*Rate of evolution of reproductive isolation*

Speed of reproductive isolation is only one contributor to speciation rate, and need not be the rate limiting step (Rabosky and Matute 2013; Price et al. 2014). Nevertheless, rate of evolution of reproductive isolation can limit truly rapid buildup. For example, if reproductive isolation always took 2 million years to evolve (the approximate tree of life average (Hedges et al. 2015), lakes Victoria and Malawi, which are much younger than 2 Ma, would presently contain very few species of African cichlid fish.

In the only comparative study of reproductive isolation and latitude, Yukilevich (2013) found that behavioral (premating) isolation evolved between sister taxa of *Drosophila* at virtually identical rates in the tropics and extratropics (subtropical and temperate zones). In contrast, hybrid sterility, which accumulates more slowly than premating isolation, evolved more than twice as rapidly in the tropics as the extratropics. Additional studies have compared rates of divergence of avian song and plumage, which are involved in mate choice and species recognition though are not direct measures of reproductive isolation. The rate of divergence between sister species of passerines in song frequency (pitch) and syllable diversity increased from the equator toward the poles (Weir and Wheatcroft 2010; Weir et al. 2012). Song syllable diversity increased with latitude in oscine passerines, but not suboscines. Similarly, the rate of divergence of plumage coloration was faster in the temperate zone than in the tropics (Martin et al. 2010). These patterns are most consistent with higher recent speciation rates at temperate latitudes than in the tropics, at least in birds.

Speciation rates might also vary latitudinally by metapopulation dynamic processes. Weir and Price (2011) showed that the rate of establishment of secondary sympatry between diverging populations of breeding birds was faster in the temperate zone than in the tropics. They attributed this pattern in part to faster evolution of premating isolation at high latitudes, which permitted coexistence sooner, but also to slower geographic range evolution and lower dispersal in the tropics, which they attributed to low ecological opportunity. Within zones, sympatry is associated with a faster rate of divergence in plumage and song between sister species of birds (Martin et al. 2015).

*Conclusions*

Studies to date suggest that recent diversification rates are not lower in the temperate zone than the tropics . Recent speciation rates might be highest in the temperate zone, at least in some taxa, which in birds and mammals appears to be associated with faster rates of evolution of reproductive isolation at temperate latitudes. In combination, the spatial patterns of speciation and diversification imply that recent extinction rates have also been higher in the temperate zone than in the tropics (Weir and Schluter 2007). However, it is not yet clear whether the signal of greater recent extinction in the temperate zone results from a Pleistocene episode or chronically higher baseline extinction.

These conclusions about recent speciation, diversification and latitude depart from the conventional wisdom and the results from multiple studies suggesting higher diversification in the tropics than the temperate zone (Mittelbach et al. 2007). Some of the disagreement between different studies might be accounted for by the time frame over which rates have been estimated.

Although the pattern is noisy, rates based on studies analyzing whole clades having deep roots in time tend to find higher speciation and diversification rates in the tropics, on average, than in the temperate zone. This contrasts with the estimates from analyses of younger clades. A reasonable interpretation of the trends is that the latitudinal gradient in speciation and diversification rate has changed over time, becoming flatter and perhaps even reversing toward the present.

The trend in diversification rate with clade age is predicted by the age and area hypothesis, but not the other two hypotheses. The earliest analyzed nodes in many taxa used to estimate speciation and diversification rates extend well into the Eocene and before, into a long period in which the world was mainly tropical. As the earth cooled, speciation and diversification rates might have increased in the seasonal temperate zone as it expanded in area. Faster modern rates of speciation in the temperate zone than in the tropics, associated with faster evolution of reproductive isolation in some taxa, are not predicted by area, however, since area is roughly similar between the temperate zone and the tropics.

More rapid recent speciation in the temperate zone than the tropics might instead be a *consequence* of the latitudinal gradient in species diversity (Weir and Price 2011), elevating ecological opportunity at high latitudes. This interpretation implies that ecological opportunity plays some role in the evolution of the latitudinal gradient. Ecological opportunity is predicted to be dynamic and should change as diversity itself changes, although there are conflicting expectations as to the direction of such change. We should expect ecological opportunity to decline if diversity buildup results in more and more species competing for a finite set of resources. On the other hand, species are themselves resources for other species, and we might therefore expect ecological opportunity to rise as diversity builds (Whittaker 1977; Schemske 2009). The data from birds, fish and mammals suggest that speciation rates increase with increasing latitude in association with reduced species diversity. This suggests that ecological opportunity might well be lower and not higher when more species are present, at least in these taxa.

**Speciation and ecological opportunity in high latitude fishes**

These broad scale patterns, while they suggest roles for geographic area and ecological opportunity in speciation rates, say little about how it happens. By themselves, the comparative data do not provide much insight into underlying mechanisms. Can detailed studies of the processes of speciation help us to evaluate the causes of observed associations between latitude and speciation? Evaluating the effects of area on speciation processes would be challenging because of the geographic scale of study required, and especially the difficulty of measuring effects on the metapopulation dynamic processes of speciation. Studies of the effects of temperature on mutation rate, generation time, and rate of evolution of reproductive isolation are feasible but have not yet been attempted. The role of ecological opportunity can be tested by measuring its presumed effects on performance and fitness within diverging populations. Under the hypothesis of ecological opportunity, adaptive peaks in performance and fitness should reflect underlying resources that in turn contribute to components of reproductive isolation and to metapopulation processes such as population persistence. To illustrate, I summarize below recent studies of the performance consequences of niche use on reproductive isolation in threespine stickleback inhabiting depauperate postglacial lakes. A single system can provide only a hint into the forces thought to influence a global pattern, but the processes discovered might apply much more broadly.

*Fast speciation in postglacial fishes*

Large tracts of the northern hemisphere were covered by ice as recently as 13,000 years ago (Clague and James 2002). The advance of these glaciers defaunated large regions that remain relatively depauperate today in many taxa. Described sister species of fish currently inhabiting North American lakes of previously glaciated areas have younger divergence times on average than species further south (Bernatchez and Wilson 1998), suggesting more recent speciation events. Many postglacial lakes additionally contain ecologically differentiated and reproductively isolated pairs of species not yet formally described. Examples are known from stickleback, smelt, whitefish, char, and trout (Schluter and McPhail 1992; Taylor 1999; Vonlanthen et al. 2012). These represent some of the youngest species on earth of any taxon, indicating frequent rapid speciation in postglacial lakes and rivers.

*Fast speciation in threespine stickleback*

To understand the evolutionary processes underlying fast speciation in postglacial fishes, we have studied the sympatric pairs of threespine stickleback found in 8 lakes of 5 small drainages on islands of SW British Columbia (Gow et al. 2008). The lakes formed at the end of the Pleistocene when submerged coastal lands rebounded after the weight of ice was removed. The marine threespine stickleback, *Gasterosteus aculeatus*, colonized virtually every new lake while it was accessible from the sea. The few lakes presently containing two stickleback species were apparently colonized twice (Taylor and McPhail 2000). Each pair consists of a limnetic ecotype that feeds on zooplankton in the open water zone and a benthic ecotype that preys upon invertebrates attached to vegetation in the littoral zone or in sediments of the lake. Sympatric species have diverged in multiple traits that increase feeding performance on lake resources. Benthics have hypertrophied epaxial musculature and jaw dimensions that enable high suction performance, advantageous for dislodging invertebrates from the vegetation and sediments (McGee et al. 2013). In contrast, the levers of the limnetic jaw provide rapid jaw opening and high jaw protrusion, aiding capture of evasive copepods in the open water (McGee et al. 2013). Solitary stickleback in otherwise similar lakes are intermediate in phenotype and ecology, suggesting ecological character displacement between the sympatric species and selection against intermediate phenotypes (Schluter and McPhail 1992; Schluter 1994; Schluter 2003). Hybrids between sympatric species are viable and fertile (Hatfield and Schluter 1999). Behavioral (premating) reproductive isolation is strong, but limited gene flow is ongoing (Gow et al. 2006).

A role for ecological opportunity on species persistence is suggested by the distribution of stickleback species pairs, which occur only in the most depauperate lakes in the region. Lakes having species pairs contain only stickleback and one other fish species, cutthroat trout (a predator), whereas the vast majority of lakes in the region contain additional fish species, notably prickly sculpin, *Cottus asper*. The sculpin is both a predator and a competitor of stickleback (Vamosi 2003). When sculpin are present, single-species populations of stickleback exhibit a shift toward a more limnetic-like phenotype (Ingram et al. 2012). This contrasts with single-species stickleback populations occurring without sculpin, which are intermediate in phenotype between benthics and limnetics. Thus, in addition to preying upon stickleback, the sculpin appears to usurp the role of the benthic ecotype. These patterns suggest that the establishment and persistence of a pair of stickleback species in a lake is facilitated by the absence of sculpin and other fish (Vamosi 2003).

*Ecological opportunity and the evolution of postzygotic isolation*

Experiments in ponds also suggest that ecological opportunity contributed to the evolution of reproductive isolation. To demonstrate, I focus on just one component of reproductive isolation: the fitness of hybrids between sympatric species (postzygotic isolation). Our initial ideas on the processes leading to the evolution of reduced hybrid fitness were based on a conceptual model similar to that shown in Figure 1. Under this view, hybrids between diverging populations are at a disadvantage because they are intermediate in phenotype and thus poorly adapted to the two ecological niches exploited by the parental species. The reduction in hybrid fitness is expected to worsen as the parental forms become better adapted to their respective environments (Rice and Hostert 1993) and under competition. In support, field enclosure experiments in which fish were confined to one habitat or the other revealed that growth performance of F1 hybrids is reduced in both habitats compared to the parental species, whereas backcross hybrids perform better in the habitat of the parental species forming the majority of their genetic makeup (Schluter 1995; Rundle 2002). In contrast, we find little evidence of reduced growth performance in the lab (Hatfield and Schluter 1999).

To investigate the impacts of resources and competition on a range of intermediate phenotypes, we measured feeding performance of subadult F2 hybrids between limnetic and benthic stickleback in an experimental pond that contained both parental habitats at a small scale (Arnegard et al. 2014). We used stable isotopes of nitrogen and carbon in muscle tissues, which reflect nitrogen and carbon in diets, to quantify niche use by individual F2 hybrids (Figure 4). Stable isotopes of carbon and nitrogen are known to differ between wild limnetics and benthics, with carbon ratio associated with littoral-pelagic habitat and diet differences and nitrogen ratio associated with differences in trophic level (Matthews et al. 2010). In a plot of nitrogen ratio against carbon ratio, limnetics occupy the upper left portion of the isotope plane (high nitrogen ratio, low carbon ratio) and benthics occupy the lower right portion (low nitrogen ratio and high carbon ratio). F2 hybrids showed wide segregation variance in isotope ratios between these two extremes (Figure 4). Strikingly, mean body size of hybrids was highest in those F2 individuals that were either the most limnetic-like in diet and niche use (upper left) or the most benthic like (lower right). Larger body size implies higher feeding performance and growth rate of individual fish. These extreme F2's were also the individuals with the greatest number of limnetic alleles, or benthic alleles, respectively, at genomic regions responsible for the species differences in phenotype. F2 hybrids intermediate in phenotype and genotype had lower feeding performance, as measured by size (Arnegard et al. 2014).

These results thus support the idea that speciation in these stickleback involved adaptation to distinct adaptive peaks determined by underlying resources. So far, they are also consistent with the simple ecological model in which hybrid fitness is a consequence of an intermediate phenotype (Figure 1). However, the study exposed an important component of hybrid performance that, if general, would significantly speed the evolution of postzygotic isolation. The result came from measurements of a group of F2 hybrids having a stable isotope signature in the lower left of the isotope plane, deviating from the major limnetic-benthic axis of niche variation (Figure 4). These F2's had the lowest mean body size of all F2 hybrid classes and an unusual phenotype (Arnegard et al. 2014). They were found to be extremely limnetic-like in some traits, and extremely benthic-like in other traits, especially in their jaws. Our interpretation is that the low performance of these hybrids resulted from the exceptionally high trait mismatch, which severely compromised the ability of individuals to feed efficiently in either of the habitats on which the parental species are specialized. Indeed, their diet in the experiment was unconventional (surface collembolans), which allowed us to detect them by their unique isotope signatures (Figure 4).

The insight from this work is that ecological opportunity, manifested as a pair of adaptive peaks corresponding to the two main lake habitats, can lead to the rapid evolution of reduced hybrid fitness by a mismatch between traits inherited from parental ecotypes adapting to the contrasting ecological environments. The mismatch causes hybrid fitness to be lower than expected based simply on their intermediate phenotype. This effect of trait mismatch on hybrid fitness was anticipated by theory (Nosil et al. 2009; Chevin et al. 2014), and can be thought of as analogous to the effects of interactions between genes known to underlie intrinsic hybrid incompatibilities (Presgraves et al. 2003). Hybrid phenotype mismatch is different from intrinsic hybrid breakdown because it is environment-dependent and expected to be most severe in the wild. Trait mismatch was not previously identified in lab studies and presumably would not be much of a hindrance there. This project is also investigating other mechanisms by which adaptation to contrasting resources has facilitated rapid speciation in threespine stickleback, including genetic linkage between divergent phenotypic traits and mate preferences (Conte and Schluter 2013), and a ready supply of standing genetic variation (Schluter and Conte 2009).

**Discussion**

The main cause of the latitudinal gradient in species diversity has eluded scientific consensus, with multiple explanations receiving support in the literature. Here I focused on three general evolutionary hypotheses for the gradient that make testable predictions about speciation rates and processes: age and area, temperature and metabolic rate, and ecological opportunity. These three general ideas essentially distill multiple evolutionary explanations for the gradient into those attributing higher tropical diversity to higher speciation rates resulting from "more area", "more mutations", and "more niches". Of these, the hypothesis of ecological opportunity further predicts that niche availability should change as species diversity builds, modifying diversification and speciation rates as a consequence. The amount of data available to evaluate the alternative predictions is neither large nor taxonomically representative, and trends identified here need to be confirmed with a larger and broader sample of studies. Tentative conclusions about speciation, latitude and the biodiversity gradient are nevertheless possible.

First, speciation and diversification rates are similar between latitudes, especially in studies of young clades. Current speciation rates might even be higher in the temperate zone than the tropics, according to some studies of taxa whose oldest analyzed dates lies within the past few million years. Second, there is an apparent trend in time: the older the deepest analyzed nodes of the clades under study, the more likely it is that rates of speciation and diversification are found to be highest in the tropics. These patterns are consistent with the age and area hypothesis, which is the only hypothesis of the three to predict a trend in average speciation and diversification rates over time. The data on which this pattern is based are noisy and come from a small and heterogeneous set of studies. Moreover, it has not yet been established whether the timing of changes in speciation and diversification rates apparent in the trend with time coincides with the timing of historical global climate changes and the development of the season temperate zone.

Age and area are not sufficient to explain faster recent speciation in the temperate zone than in the tropics in some taxa, however. A reasonable interpretation of apparent recent bursts of speciation in the temperate zone is that expansion of the zone, including the re-opening of vast areas following de-glaciation, generated high ecological opportunity that has boosted speciation rates. If so, then present-day relationships between speciation and latitude are affected by low species diversity at high latitudes. In other words, rather than demonstrating how differences in speciation rate predict the latitudinal biodiversity gradient, present-day patterns of speciation with latitude are partly a consequence of the latitudinal gradient in species diversity. These bursts of speciation in depauperate post-glacial environments are like other episodes of rapid speciation and diversification in the history of life, which are typically associated with geological upheaval, mass extinction, colonization of new environments, and the acquisition of key evolutionary innovations. Combined with results from detailed studies of speciation mechanisms in fishes of depauperate lakes, the findings suggest that ecological opportunity has contributed to the evolution of the latitudinal biodiversity gradient. However, since ecological opportunity is dynamic and likely to change as species diversity itself changes, it may be inappropriate to use contemporary patterns of speciation and diversification to estimate rates in the more distant past.

Most estimates of speciation and diversification included in the present survey rely upon phylogenetic methods whose accuracy in such applications has not been confirmed. Analyses based on measurements of the reconstructed number of lineages through time, and those based on measured associations between geographical location and rates of diversification and speciation (BiSSE, GeoSSE; Table A1), assume that within latitudes the rates are constant through time and along all branches of the phylogenetic trees. The assumption of rate constancy through time is contradicted by apparent changes through history in the ratio of temperate and tropical speciation and diversification rates (Fig. 2, 3). Heterogeneity of rates across tree branches within latitudes is often striking (Jetz et al. 2012; Rabosky et al. 2015). Rate heterogeneity can bias estimates of extinction rates, and can bias estimated associations between traits (here, latitude) and speciation and diversification (Rabosky 2010; Rabosky and Goldberg 2015). Some of these concerns are addressed by methods that accommodate rate heterogeneity, such as BAMM (Rabosky et al. 2015), but they assume discrete partitions in rates and have limited power to detect rate heterogeneity (Rabosky and Goldberg 2015). The phylogenetic methods have also recently been criticized because they greatly underestimate the sampling error of estimates of diversification and speciation (Rabosky and Goldberg 2015). Noisy rate estimates would not affect the results presented here, other than make trends more difficult to detect. However, biases resulting from violations of model assumptions are likely. Perhaps the biases affect estimates of diversification and speciation rates similarly in the tropics and temperate zone, in which case the ratio of rates between the two zones would be less sensitive.

Comparisons made of the same clades at different depths in the phylogeny suggest that the trends with time are robust, at least in some cases. For example, Ricklefs (2006) showed that primarily tropical clades of New World passerine birds are more diverse than temperate clades of the same age, implying that passerine diversification rates over time were highest in the tropics on average. The tribes and families used in that study have an oldest age of about 80 ma (Table A1), which predates the Eocene and represents a point in relatively deep time. In contrast, the analysis of Jetz et al. (2012) of the same New World passerines found no clear difference in diversification rates between latitudes. Although their phylogeny included the same clades and hence time frame, Jetz et al. (2012) used a species-level metric that estimates diversification rate near the tips of the phylogeny, and so measures recent rates. The same species-level metric also found no latitudinal gradient in recent diversification rates in mammals (Belmaker and Jetz 2015), in contrast to the likelihood-based analyses (using GeoSSE) of the whole tree of mammals, which extends back more than 200 my and showed highest diversification rates in the tropics (Rolland et al. 2014). These comparisons in birds and mammals are consistent with the finding that recent rates of diversification are similar between the tropics and temperate zone, whereas the rates were higher in the tropics in the past.

A major assumption of the present study is that speciation is a rate-limiting step in the accumulation of diversity. An alternative point of view is that the latitudinal diversity gradient is the result of resource limitation and saturation, which allows more species to coexist in the tropics than the temperate zone (Marshall 2007; Valentine et al. 2008). In this case, rates of speciation and diversification are not the main cause of the latitudinal gradient, and estimates based on the assumption of rate constancy will give misleading results (Rabosky 2009). The concept of saturation is challenging to apply to species and other taxa whose traits continue to evolve. A hypothetical saturation point for species richness would be altered after each evolutionary innovation and with changes in dispersal and mating characteristics of component taxa. The concept of a diversity limit might nevertheless be useful when applied to a specific taxon over a predefined time period in which the distribution of phenotypes isn't changing much. The ceiling would be set, perhaps, by the total density of individuals that can be supported on the resources available. However, density of individuals shows no straightforward latitudinal gradient within taxa (Tilman and Pacala 1993; Enquist and Niklas 2001; Currie et al. 2004). In addition, despite evidence of diversity-dependent speciation from fossil and biogeographic data, there is little indication that species richness has reached saturation at all latitudes (Crame 2001; Cornell 2013; Rabosky and Hurlbert 2015). The success of invasive species, both human-caused (Stohlgren et al. 2005; Sax et al. 2007) and natural (Ricklefs 2002; Jablonski et al. 2006; Kennedy et al. 2014; Ricklefs and Jønsson 2014) argues against saturation as a general condition. Variable degrees of convergence of species diversity of taxa between similar sites in different parts of the world also argues against the ubiquity of saturation (Schluter and Ricklefs 1993).

A related idea is that species diversity in all global regions is presently at or near an equilibrium influenced by available resources, where rates of speciation, colonization, and extinction are balanced (Rosenzweig 1995; Rabosky and Hurlbert 2015). This is a weaker claim than saturation because equilibrium requires only that input rates balance output rates, in a continental analogue of the island biogeographic equilibrium. In this case, a change of input rates changes the equilibrium, and so the idea can accommodate the assumption made here that speciation rate is a limiting step in the evolution of diversity. However, there is currently little evidence that diversity has everywhere reached stable equilibrium (Cornell 2013; Harmon and Harrison 2015).

The success of natural and human-caused invasions to continental regions, often without fatal consequences for endemic species, suggests that ecological niches are frequently underutilized (Sax et al. 2007). This implies in turn that adaptation is constrained, such that the evolution of a wide diversity of phenotypes able to exploit many ecological niches requires a great deal of time. Otherwise, it is difficult to explain why native taxa had not already evolved and diversified to exploit the ecological niches now occupied by invaders. One possible explanation is that niches colonized by invasive species are new, freshly created by recent climate change or disturbance. This is especially likely in the case of invasive species living in human-modified landscapes. However, longer-term adaptive constraints are probably ubiquitous. Fitness landscapes are complex and multi-modal, and evolution in a population to occupy a new adaptive peak only occurs when all intermediate stages are advantageous (e.g., Coyne et al. (1997)). The transition between adaptive peaks separated by an adaptive valley at a given location might thus often require emigration to, and adaptation in, a sequence of alternative environments having different optima that circumvent the valley, followed by back-colonization. In this way an immigrant can often do what an endemic lineage cannot, at least not in time available.

One of the goals of this essay was to ask whether detailed studies of speciation mechanisms might help to determine the causes of major biodiversity gradients. Tests of mechanisms underlying the latitudinal gradient have chiefly relied on comparative data on rates. However, general hypotheses for the gradient make predictions about mechanisms of speciation along the gradient and how these processes lead to speciation rate variation. For example, the age and area hypothesis predicts that greater area should promote speciation via the metapopulation dynamics of the process more than via the speed of evolution of reproductive isolation. The temperature and metabolism hypothesis predicts the opposite, that is, the increase in rate of divergence at higher temperatures should occur mainly by higher production of advantageous mutations. The hypotheses for the gradient might also make different predictions concerning which mode of speciation, ecological or mutation-order, should be most affected. For example, higher temperature was one explanation proposed for why hybrid male sterility (whose rate of evolution is likely mutation-limited) evolved faster in *Drosophila* in the tropics than the extratropics (Yukilevich 2013). However temperature was not the only explanation provided for the pattern. Testing these predictions awaits more comprehensive studies of speciation mechanisms across latitudes.

Here, I have used studies of speciation in stickleback in high-latitude lakes to address a prediction of the hypothesis of ecological opportunity. I concentrated on factors contributing to the evolution of just one component of postzygotic isolation, namely the reduced performance of hybrids owing to an intermediate and mismatched phenotype. These effects seem clearly to be evolved outcomes of adaptation to contrasting ecological niches. Multiple traits (and genes) are involved in divergence between ecotypes adapting to alternative littoral and zooplankton resources. The work suggests that one way in which ecological opportunity can speed the evolution of reproductive isolation is by the increasing mismatch of hybrid phenotypes as parental species continue to adapt to contrasting ecological niches. These results show that at the very least the mechanisms at the heart of the hypothesis of ecological opportunity are at work, and they may help to explain elevated speciation rates at temperate latitudes in the recent past, We have obtained some, but unfortunately only limited, insight into the contribution of depauperate environments to the metapopulation dynamics of the speciation process, and this is true more generally.

What role might ecological opportunity have played in the evolution of the modern latitudinal gradient? The available results on speciation rate suggest that ecological opportunity is presently highest in the temperate zone, although the number of taxa investigated so far is limited. Perhaps the effect is real but will be short-lived, with ecological opportunity for the most part being highest in the tropics, and playing a major role in the evolution of the modern latitudinal diversity gradient. For example in fishes, no temperate zone radiation, including lineages in old, large Lake Baikal, can match the diversity or sustained high diversification and speciation rates of the African cichlids. Other examples of speciation associated with high ecological opportunity are known from the tropics, such as Bahamian pupfish (Martin and Wainwright 2013). On the other hand, the available time and number of lineages in the temperate zone overall are reduced, and no fair comparison of diversification and speciation rates in tropical and temperate fishes has yet been conducted (none are included in Table A1). In general, latitudinal gradients in phenotypes suggest a greater variety of resources and thus more niches in the tropics than in the temperate zone. A proper test of whether ecological opportunity has contributed to the latitudinal gradient via enhanced speciation rate would require speciation records over a stretch of time beginning from multiple regions of equal area and equal diversity in the tropics and temperate zone. While this experiment is not possible, a surrogate study might compare recent rates of speciation and diversification in similar, newly originated environments at a range of latitudes in the recent past, with similar resource diversities and abundances and colonized by a similar number of phylogenetically comparable lineages (see Wagner et al. (2012) for a comparative study of cichlid fishes along these lines, though without all necessary controls). A win for the tropics under this scenario would imply that ecological opportunity alone could produce a latitudinal gradient, although it would not detract from the conclusion that our modern latitudinal gradient has been influenced by age and area of the tropical and temperate zones.

One might equivalently argue that higher species diversity and lower per lineage speciation rates at low latitudes is recent and is obscuring the positive long-term effects of higher temperature and metabolic rate on speciation rates in the tropics. Modification of the original temperature and metabolic rate hypothesis to include diversity-dependent feedbacks predicts that the effects of temperature on speciation (and extinction) rates diminish as species diversity builds (Stegen et al. 2009). This idea might also be tested by comparing current rates of speciation and diversification between lineages at different latitudes while controlling for species diversity, resources and the number of phylogenetically comparable lineages.

Although the age and area hypothesis predicts the apparent temporal trend in the ratio of temperate and tropical speciation and diversification rates, the hypothesis leaves open the question of why the temperate zone has not already caught up with the tropics in species diversity. Since the areal extent of the temperate zone is roughly similar to that of the tropics, why is species diversity not now much more similar? Tens of millions of years of global cooling seems a sufficiently long time if one considers the lightening speed at which speciation is known sometimes to occur, and considering that species turnover has been occurring all the while. However, the time frame for the present-day latitudinal gradient might only be about 15 million years, the period of largely continuous global cooling since the warm mid-Miocene optimum. The boreal biome is even younger (Fine and Ree 2006). As well, area of the temperate zone was greatly restricted during the Pleistocene. Finally, although speciation can sometimes occur at great speeds, the average duration of speciation is about 2 Ma (Hedges et al. 2015), which is not an insignificant portion of the time frame over which the modern latitudinal gradient developed. In general, speciation and diversification rates are likely to be highly heterogeneous among clades. For example in birds, colonization of the tropics from the temperate zone by the Thraupidae (tanagers) produced 400 new species over the past 13 Ma (Kennedy et al. 2014). However, many other clades that colonized the tropics have not diversified more than their temperate zone counterparts, and over all birds there is no latitudinal gradient in the occurrence of clades having exceptional (or unexceptional) speciation and diversification rates (Jetz et al. 2012; Rabosky et al. 2015).

Moreover, speciation continues relentlessly at all latitudes, even as the temperate zone might lately have been experiencing a faster per-lineage rate of speciation than the tropics. Temperate zone speciation rates might not be drastically higher than those in the tropics, and per-lineage diversification rates are not much different between latitudes. Thus, the main reason why the temperate zone has not caught up with the tropics might simply be that the speciation differential is not large, and there are so many more lineages in the tropics to begin with. A greater number of lineages in the tropics ensure that the total number of new species produced per unit time will be highest there. Quite possibly, history has given the tropics a large head start, and recent latitudinal gradients in evolutionary rates have been insufficient to overcome this "tropical evolutionary momentum" in the time available. For this reason, the latitudinal diversity gradient persists.

## Acknowledgements

I grateful to Jedediah Brodie, Greg Crutsinger, Darren Irwin, Chase Mendenhall, Arne Mooers, Mary O'Connor, Walter Jetz, Trevor Price, Loren Rieseberg, Doug Robinson, Corey Tarwater, Jason Weir, and Mike Whitlock for discussion. The manuscript was much improved by input from many of these same people plus Allen Herre, Sara Miller, Gary Mittelbach, Chris Muir, Sally Otto, Dan Rabosky, Diana Rennison, Bob Ricklefs, Seth Rudman, Kieran Samuk, and Doug Schemske. I thank the staff at Las Cruces field station of the Organization of Tropical Studies, and Allen Herre and the Smithsonian Tropical Research Institute for hosting me while I wrote parts of this paper. My work is supported by the Natural Sciences and Engineering Research Council of Canada, the National Institutes of Health, the Canada Foundation for Innovation, and the Canada Research Chairs.

Figure 1. A visual depiction of speciation on a fitness landscape under ecological opportunity, in which populations adapt to alternative ecological niches. Divergent selection leads ultimately to the evolution of reproductive isolation as a by-product. In an alternative mechanism also involving ecological opportunity, spatially separate populations adapt to similar ecological niches but by chance acquire and fix alternative mutations that cause reproductive incompatibilities.

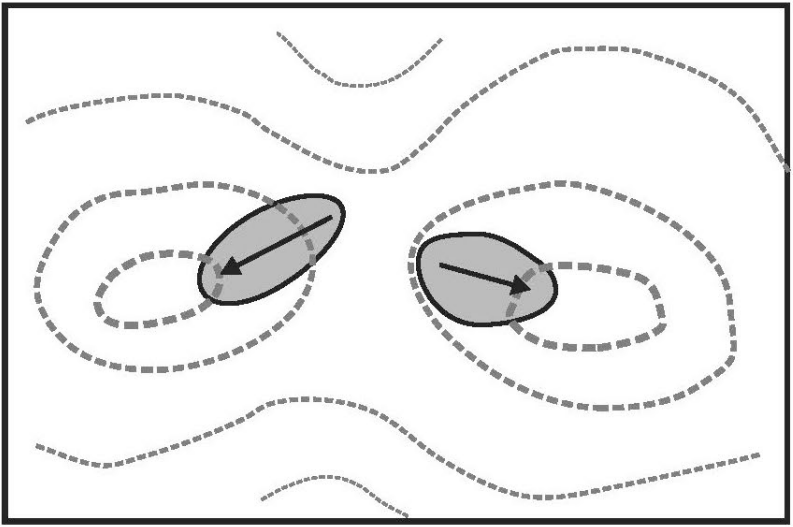

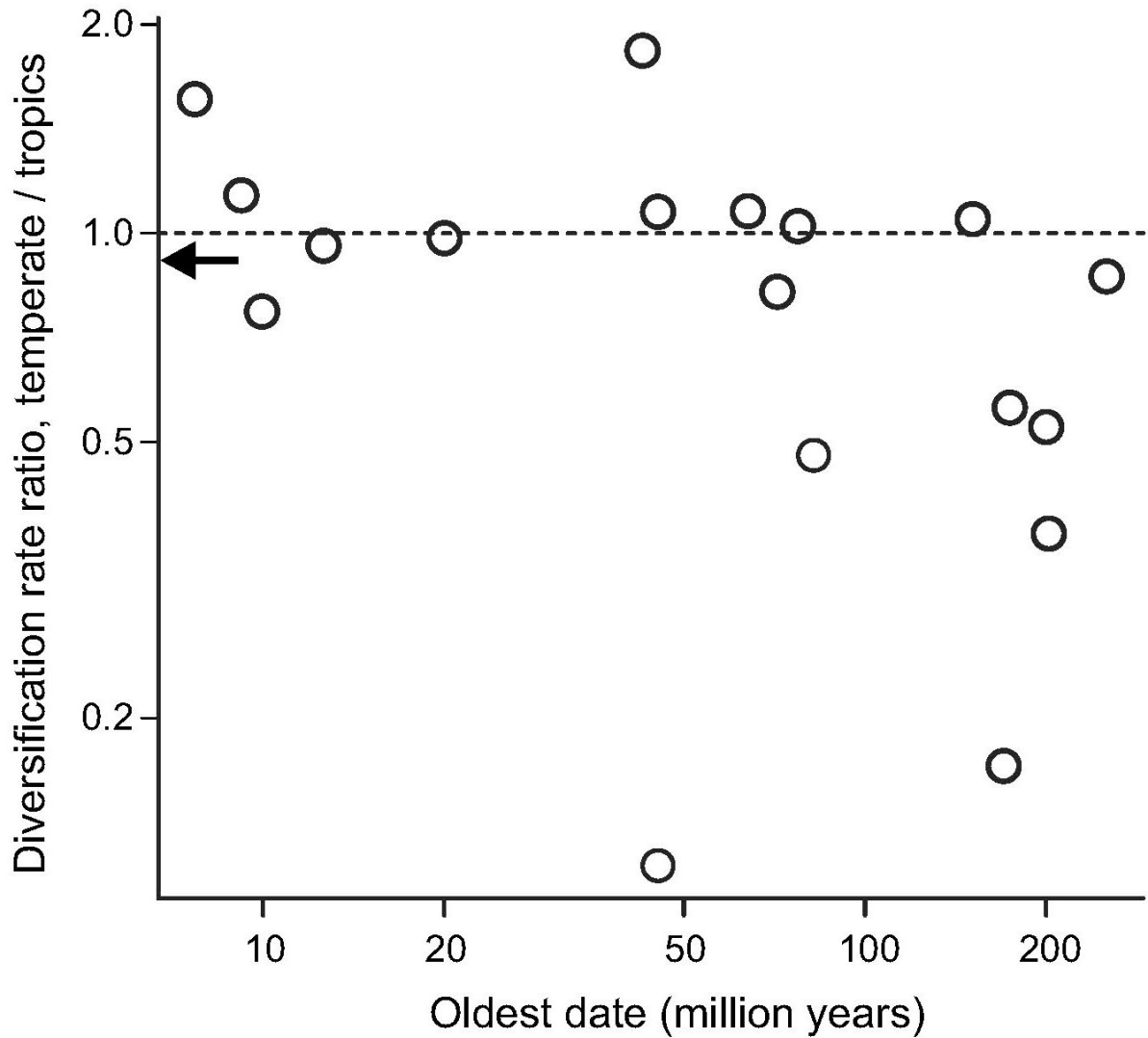

Figure 2. The ratio of estimated diversification rates in the tropics and the temperate zone in relation to the time span analyzed, as measured by the oldest date represented. The arrow indicates the median ratio. Data are from Table A1.

Figure 3. The ratio of estimated speciation rates in the tropics and the temperate zone in relation to the time span analyzed, measured by the oldest date represented. The arrow indicates the median ratio. Data are from Table A1.

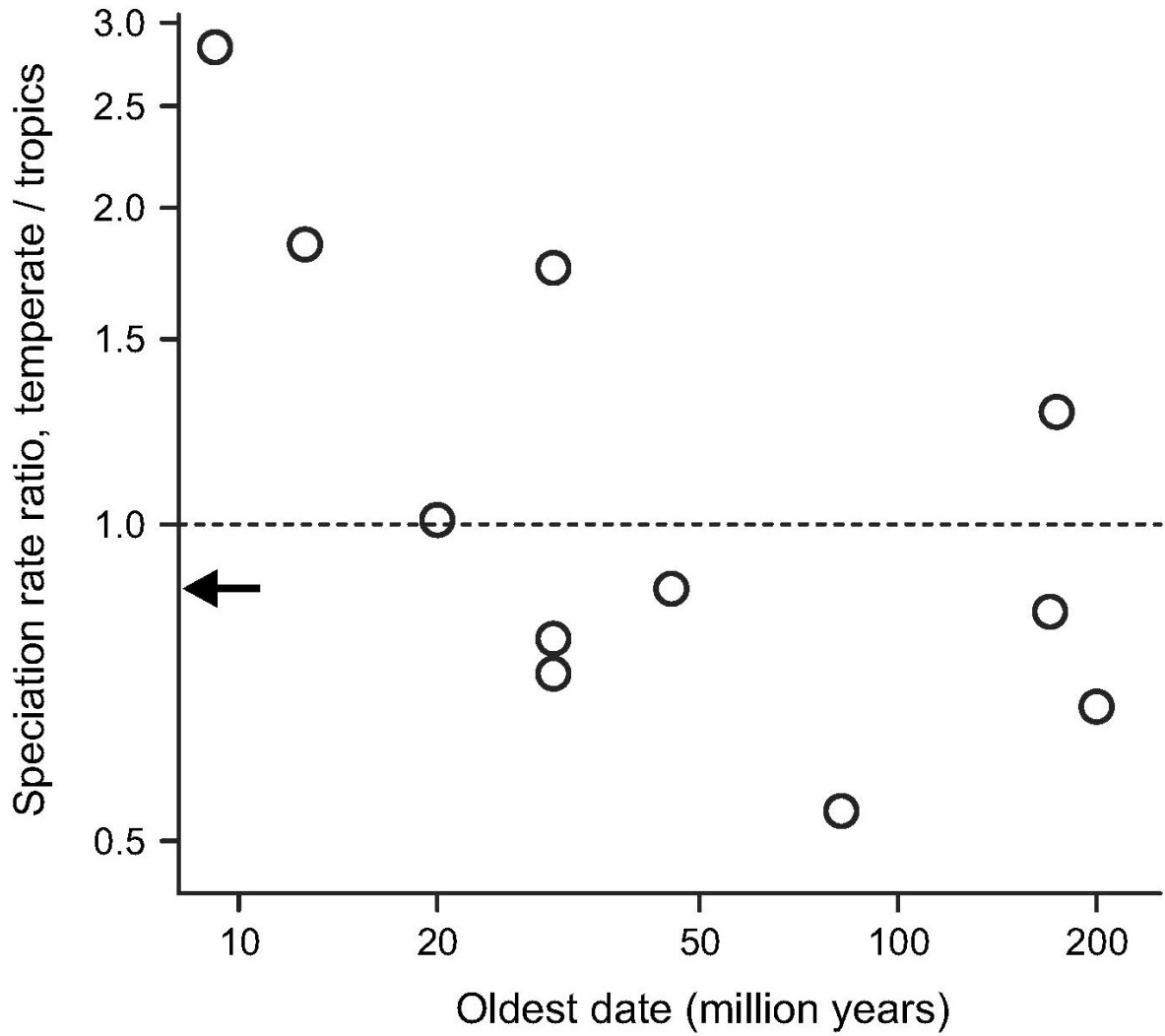

Figure 4. Carbon (C) and nitrogen (N) stable isotope composition of individual F2 hybrids (open circles) between limnetic and benthic stickleback species in an experimental pond. Individuals in the upper left have a more limnetic-like diet and isotope signature, whereas those in the lower right are more benthic-like. Contours estimate mean body size (standard length, in mm, reflecting rate of growth) of individual fish having different carbon and nitrogen isotope signatures. Modified from Arnegard et al. (2014).

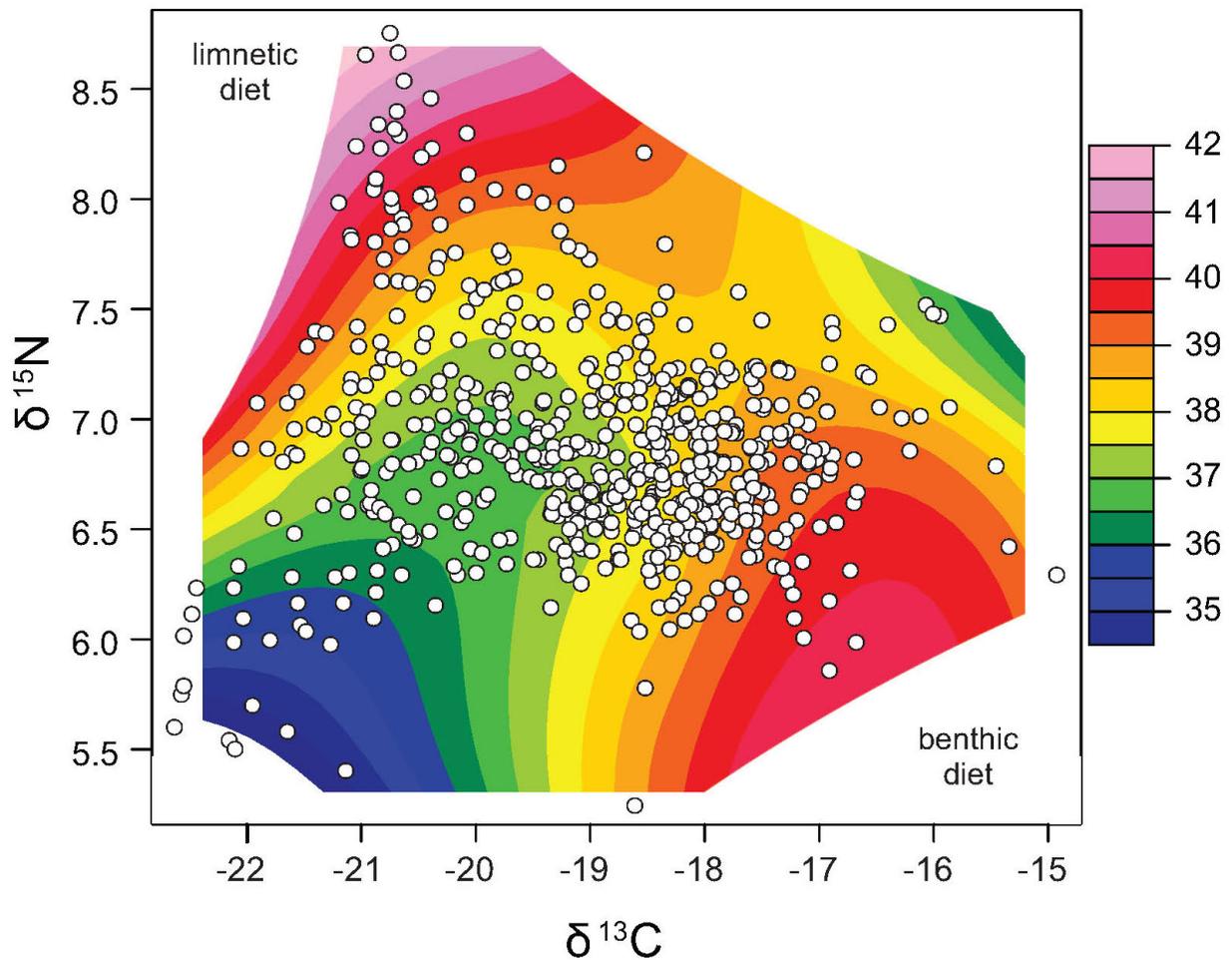

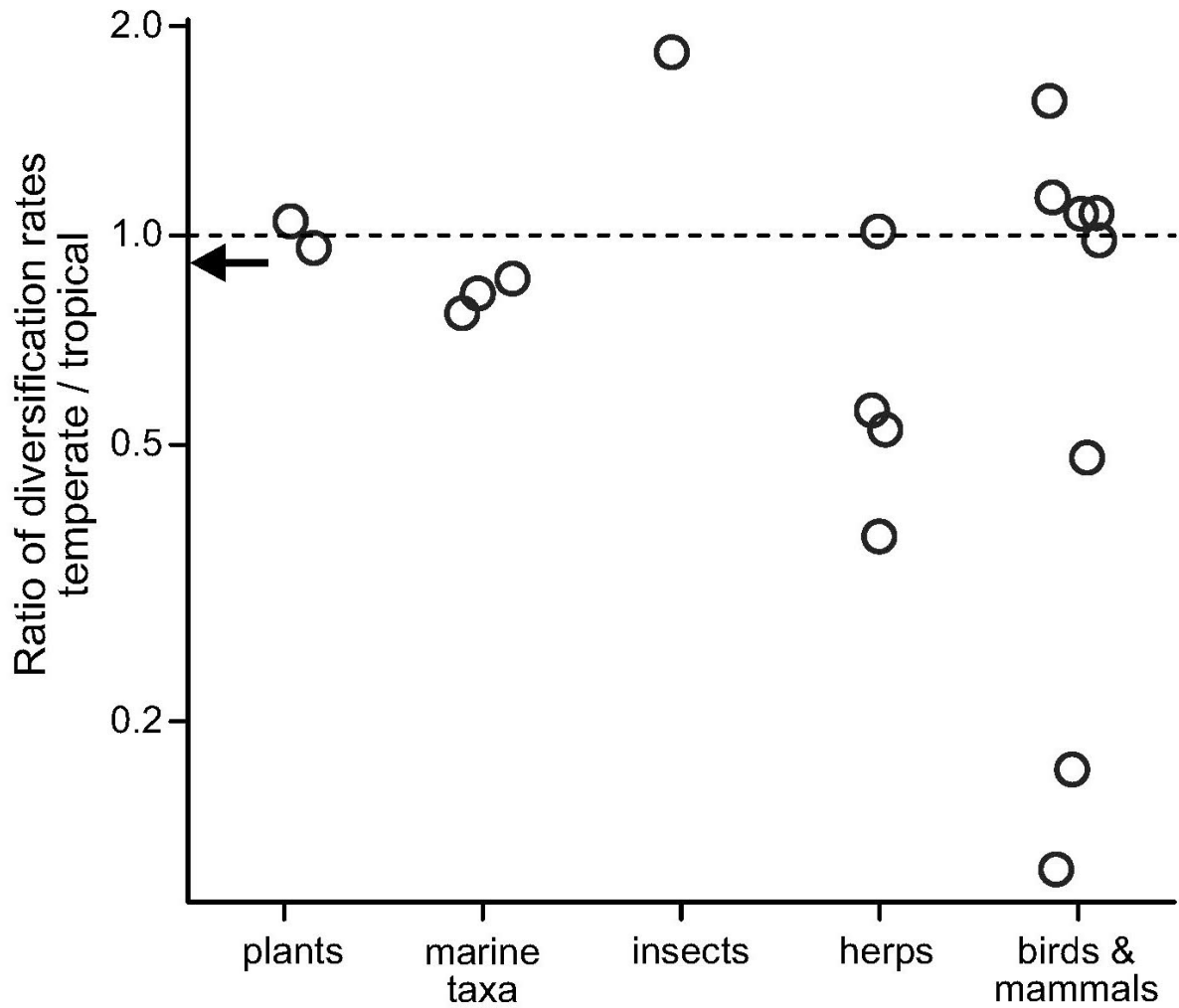

Figure A1. The ratio of estimated diversification rates in the tropics and in the temperate zone from different study taxa. The arrow indicates the median diversification rate. Data are from Table A1.

# Appendix. Estimates, taxa, and data sources.

Table A1. Estimates of per capita (per lineage) diversification and speciation rates and their sources.

| Taxon | Source | Method | Oldest date (Ma) | Region | Temperate rate | Tropical rate | Ratio Temp/Trop | Notes |
|---|---|---|---|---|---|---|---|---|
| **DIVERSIFICATION RATES** | | | | | | | | |
| Plants | Davies et al. (2004); Jansson and Davies (2008) | log($S$) ratio | 151 | Global | | | 1.048 | Calculated from sister species count (J. Davies, pers. comm. 2014)[1]. |
| Plants | Jansson et al. (2013) | log($S$) ratio | 12.6 | Global | | | 0.959 | Temperate and equatorial sister species data from their Table S1[2]. |
| Benthic foraminifera | Buzas et al. (2002) | Fossil species counts at two time points | 10.5 | N Atlantic Ocean | 0.030 | 0.040 | 0.772 | Log of Fisher's α at 1 and 10.5 Ma[3] from their Figure 1. |
| Benthic marine invertebrates | Kiessling et al. (2010) | Relative risk of fossil genus originations | 252 | Global | | | 0.867 | Calculated from the log of relative risk in lower panel of their Fig. S8[4]. |
| Marine bivalves | Krug et al. (2009) | Taxon survivorship analysis | 71.5 | Global | 0.018 | 0.022 | 0.823 | Origination rates for ~711 genera from their Fig. 3[5]. |
| Insects | Jansson et al. (2013) | log($S$) ratio | 42.7 | Global | | | 1.834 | Temperate and equatorial sister species data from their Table S1[6]. |
| Amphibians | Pyron and Wiens (2013) | GeoSSE | 200 | Global | 0.030 | 0.057 | 0.526 | Values provided in their main text[7]. |
| Hylid frogs | Wiens et al. (2011) | Magallón and Sanderson (2001) | 78 | Global | 0.063 | 0.061 | 1.027 | Data from their Table 1, which use a relative extinction of ε=0.45[8]. |
| Squamate reptiles | Pyron (2014) | GeoSSE | 174 | Global | 0.038 | 0.067 | 0.560 | Values under the full model provided in his Table 1[9]. |
| Birds | Ricklefs (2006) | Magallón and Sanderson (2001) | 82 | Western Hemisphere | 1.46 | 3.05 | 0.479 | Values from his Table 1.[10] |
| Birds | Jetz et al. (2012) | Species-level diversification | 20 | Global | 0.165 | 0.167 | 0.984 | Values from their Figure 4[11]. |
| Birds | Cardillo et al. (2005) | log($S$)/$t$ | 45.3 | Western Hemisphere | 0.011 | 0.093 | 0.123 | Data from their Fig. 2a[12]. |
| Passerine birds | Cardillo (1999) | log($S$) ratio | 63.9 | Global | | | 1.076 | Based on data from his Table 1[13]. |
| Birds & mammals | Weir and Schluter (2007) | Age distribution of sister species | 9.2 | Western Hemisphere | 0.152 | 0.134 | 1.134 | Values provided in text[14]. |
| Mammals | Jansson et al. (2013) | log($S$) ratio | 7.7 | Global | | | 1.561 | Based on temperate and equatorial data from their Table S1[15]. |
| Mammals | Soria-Carrasco and Castresana (2012) | Magallón and Sanderson (2001) | 45.3 | Global | 0.176 | 0.164 | 1.074 | Data from their Table S6[16]. |
| Mammals | Rolland et al. (2014) | GeoSSE | 170 | Global | 0.019 | 0.112 | 0.171 | Numbers are from their Table S1, based on the whole phylogeny[17]. |

**SPECIATION RATES**

| | | | | | | | | |
|---|---|---|---|---|---|---|---|---|
| Plants | Jansson et al. (2013) | GeoSSE | 12.6 | Global | 0.120 | 0.065 | 1.846 | Values are from their Table 1 of Appendix S2[18]. |
| Foraminifera | Allen and Gillooly (2006) | Fossil rate of first occurrences | 30 | Global oceanic | 0.148 | 0.205 | 0.721 | Based on values in their Table 3[19]. |
| Nannoplankton | Allen and Gillooly (2006) | Fossil rate of first occurrences | 30 | Global oceanic | 0.164 | 0.211 | 0.778 | Based on values in their Table 3[19]. |
| Radiolaria | Allen and Gillooly (2006) | Fossil rate of first occurrences | 30 | Global oceanic | 0.317 | 0.181 | 1.754 | Based on values in their Table 3[19]. |
| Amphibians | Pyron and Wiens (2013) | GeoSSE | 200 | Global | 0.038 | 0.057 | 0.671 | Values provided in their main text. |
| Squamate reptiles | Pyron (2014) | GeoSSE | 174 | Global | 0.086 | 0.067 | 1.279 | Values under the full model provided in his Table 1. |
| Birds | Ricklefs (2007) | Distribution of clade sizes (Bokma 2003) | 82 | Western Hemisphere | 3.160 | 5.920 | 0.534 | Based on values provided in his main text[20]. |
| Birds | Rabosky et al. (2015) | BAMM | 20 | Global | 0.138 | 0.137 | 1.010 | Values obtained from Fig. 2b[21]. |
| Birds & mammals | Weir and Schluter (2007) | Age distribution of sister species. | 9.2 | Western Hemisphere | 0.552 | 0.194 | 2.843 | Values provided in text[14]. |
| Mammals | Soria-Carrasco and Castresana (2012) | Nee et al. (1994); Morlon et al. (2010) | 45.3 | Global | 0.280 | 0.323 | 0.868 | Based on values in their Table S7[22]. |
| Mammals | Rolland et al. (2014) | GeoSSE | 170 | Global | 0.093 | 0.112 | 0.826 | Numbers are from their Table S1, based on the whole phylogeny[17]. |

Note: These rates in the tropics and temperate zone are those provided in the source publications or are rates predicted at 0° and 40° absolute latitude from regressions. Values from different studies may not be in comparable units, but the units cancel in the ratio of the temperate and tropical rates. In some cases the ratio itself was regressed against latitude if the per lineage rates were not provided. Rates were converted to per capita rates by division, if necessary. Data underlying new calculations of diversification and speciation at different latitudes from sources are deposited in the Dryad Digital Repository: http://dx.doi.org/10.5061/dryad.248vp (Schluter 2015).

    Genus origination rates (Krug et al. 2009; Kiessling et al. 2010) were classified as diversification rates rather than speciation rates because of the possibility of species extinctions during the time to origin of a genus. Oldest age refers to the age of the oldest fossil stratum in the data set, or the age of the deepest node in the set of trees analyzed. These were obtained from the source publications when provided, or from www.timetree.org otherwise (Hedges et al. 2006). The exception is the oldest date for the global analysis of birds analyzed by Jetz et al. (2012) and Rabosky et al. (2015), because they estimate diversification and speciation rate, respectively, at the tips of the avian phylogeny. I conservatively assigned these two studies an oldest date of 20 Ma, which includes virtually all transitions in the species-level metric of diversification rate in Fig. 2 of Jetz et al. (2012). To minimize extrapolation I included all known published studies whose estimates spanned at least 30° latitude or a 30° difference in latitude. To minimize redundancy, the study by Tamma and Ramakrishnan (2015) was not included because they analyzed a subset of the same mammal

tree analyzed by Rolland et al. (2014) using the same method. Similarly, Kennedy et al. (2014) reanalyze a subset of the Jetz et al. (2012) data set with similar findings. When a single study obtained multiple estimates for the same taxon using different methods I li estimates only from the most comprehensive analysis. To reduce sampling error, I excluded tropical-temperate sister-group studies, and studies regressing rates against latitude across multiple clades, if based on fewer than 10 sister groups or clades.

Fossil species counts at two time points were used to estimate per capita diversification rate assuming exponential growth. Fossil estimates of speciation rate are based on the rate of first occurrences. The simplest phylogenetic estimate of diversification fro phylogenetic trees was $\log(S)/t$, or $\log(S + 0.5)/t$ where $S$ is the number of extant species of a clade and $t$ is the age of the earliest split in the clade. Rates could not be computed for sister-group diversity data lacking ages; instead, the ratio of the log number of species $S$ of temperate and tropical sisters was regressed on difference in latitude and used to predict ratios at 0° and 40°. GeoSSE (Goldberg et al. 2011) is an extension of the BiSSE method (Maddison et al. 2007) for estimating range evolution and diversification from phylogenetic data. BAMM is a Bayesian method that quantifies variation in speciation and diversification rates along the branches of phylogenetic trees (Rabosky 2014).

---

[1] The ratio of diversification was calculated for each sister pair as $\log(S + ½)$ of the clade having the higher mean latitude divided by the same quantity in the lower-latitude clade. This ratio was then regressed against the difference in latitude between the two siste clades. To meet assumptions, a log-transformation was used to fit the line and obtain the predicted ratio at a difference in latitude of 40°, which was then back-transformed. The mean diversification ratio was 1.747.

[2] The ratio of diversification was calculated for each sister pair as $\log(S + ½)$ of the temperate clade divided by the same quantity in the tropical clade. The value shown is the mean (the median was 1.000). A GeoSSE analysis of a subset of the plant data estimated diversification to be 0 in the temperate zone and 0.039 in the tropics, yielding an infinite ratio, contradicting this sister-pair analysi

[3] Values of Fisher's α were analyzed as species counts. Diversification was calculated for the tropics as $(\log(37) - \log(18.1))/18.1$, with the denominator included to obtain a per-capita rate. The rate for the temperate zone was calculated similarly.

[4] The ratio is the relative risk of originations at tropical and extratropical latitudes, calculated as the exponential of the log relative risk, based on all the data from the Mesozoic-Cenozoic given in their Figure S8. The relative risk was then inverted so that tropical risk was in the denominator.

[5] Data on origination rates per lineage per million years for 27 provinces and latitudes were extracted from their Figure 3. Diversification rate was computed as the average of post-K/Pg and pre-K/Pg origination rates, weighted by the number of genera in each period (640 and 71, respectively). Diversification was regressed against latitude, and the regression line was used to predict diversification at 0 (tropical) and 40° (temperate) latitudes.

[6] The ratio of diversification was calculated for each sister pair as $\log(S + ½)$ of the temperate clade divided by the same quantity in the tropical clade. The value shown is the mean (the median was 1.190).

[7] Results are as stated in the main text of source, obtained from a GeoSSE analysis. The authors carried out an additional analysis in which diversification rates of 66 families (estimated using a simple birth-death model (Nee et al. 1994) and adjusted for incomplete sampling) against mean family latitude. I used this regression to predict diversification at 0 and 40° latitude, yielding a ratio of 0.36 Data were provided in their Tables S14 and S15.

[8] Mean diversification was regressed against latitude for 12 locations given in Table 1 and the regression was used to predict diversification at 0 and 40° latitude. The ratio of diversification was virtually identical when the island (West Indies) point was

dropped. Diversification was based on an estimate of relative extinction of 0.45, but the authors say that results were similar if they assume relative extinctions of 0 and 0.90 instead.

[9] Diversification rate was calculated in each zone as the difference between estimated speciation and extinction rates.

[10] Values are estimated diversification rates in North (mainly temperate) and South America (mainly tropical) assuming that relative extinction is 0.90, the middle of the range of values considered (0.80 to 0.98). Ratio of diversification is similar for lower and higher values of relative extinction (the range was 0.536 to 0.352). Using the same data, Ricklefs (2007) used the frequency distribution of clade sizes to estimate both speciation and extinction separately (Bokma 2003). Using these values instead yields a diversification ratio of 0.496.

[11] Predicted values of species-level diversification at 0° and the average of diversification at −40° and 40° were obtained directly from the nonlinear curve.

[12] Diversification rate was fitted to latitude using a logistic regression and extrapolated to predict diversification at 0° and 40° latitude.

[13] The ratio of diversification was calculated for each sister pair as $\log(S + ½)$ of the clade having the higher mean latitude divided by the same quantity in the lower-latitude clade. This ratio was then regressed against the difference in latitude between the two sister clades. To meet assumptions, a log-transformation was used to fit the line and obtain the predicted ratio at a difference in latitude of 40°, which was then back-transformed. The mean diversification ratio was 0.894.

[14] The relationship between diversification and latitude was calculated as the difference between the relationships of speciation $\lambda$ and extinction $\mu$ with absolute value of latitude: $\lambda = 0.194 + 0.00894*latitude$ and $\mu = 0.06 + 0.00848*latitude$. The diversification relationship was then used to predict rates at 0° and 40° latitude. The relationship between speciation and latitude was used to predict speciation rates at 0° and 40° latitude.

[15] The ratio of diversification was calculated for each sister pair as $\log(S + ½)$ of the temperate clade divided by the same quantity in the tropical clade. The value shown is the mean (the median was 1.00).

[16] Values are based on data extracted from their Table S6. One outlier having diversification greater than 3 was deleted. Diversification was regressed against latitude and used to predict values at 0° and 40° latitude. Values shown are based on a relative extinction of 0. The ratio of diversification was 1.013 when relative extinction of 0.9 was used instead.

[17] Values were obtained using the model having the best fit to the data. Diversification rates in the tropics and temperate zone were calculated as the differences between estimated speciation and extinction rates. A related study by Tamma and Ramakrishnan (2015) analyzed the subset of Asian mammals using the same method and obtained a ratio of diversification rates, temperate to tropics, of 0.178, and a ratio of speciation rates of 0.478.

[18] Values are from their multi-clade analysis in Table 1 of Appendix S2.

[19] Data are the rate of first occurrences of new morphospecies in the fossil record. Speciation rates from their Table 3 were converted to per capita rates per million years and regressed on the absolute value of latitude. The regression was then used to predict speciation rates at 0° and 40°.

[20] Values are based on relative time. Speciation rate estimates in North (mainly temperate) and South America (mainly tropical) are estimated from the frequency distribution of clade sizes.

[21] Temperate zone speciation rate is the average of predicted values at −40 and 40° latitude, whereas the tropical rate is that predicted at 0° latitude.

[22] Speciation rates of genera were regressed on the absolute value of latitude and used to predict rates at 0° and 40°. One outlier having speciation rate greater than 3 was deleted.